\documentclass[conference]{IEEEtran}
\usepackage{amsthm}
\usepackage[utf8]{inputenc}
\usepackage[english]{babel}
\usepackage{amsfonts}
\usepackage{verbatim}
\usepackage{mathtools, cuted}
\usepackage{adjustbox}
\usepackage{amsmath}
\usepackage{graphicx}
\usepackage[lofdepth,lotdepth]{subfig}
\usepackage{graphics}

\usepackage{tabularx}
\usepackage{caption}
\usepackage{cite}
\usepackage{lipsum}
\usepackage{color}
\usepackage[ruled,linesnumbered,vlined]{algorithm2e}
\usepackage[compact]{titlesec}
    \usepackage{setspace} 
    
\usepackage[bottom = 0.985in, top=0.7in, right=0.7in, left=0.7in]{geometry}

\def\BibTeX{{\rm B\kern-.05em{\sc i\kern-.025em b}\kern-.08em
    T\kern-.1667em\lower.7ex\hbox{E}\kern-.125emX}}

\usepackage{amssymb,amsmath,amsthm,enumerate}
\usepackage[utf8]{inputenc}
\usepackage{array}
\usepackage{graphicx}
\usepackage{caption}
\usepackage{subcaption}
\usepackage{bm}
\usepackage{amsfonts,amscd}
\usepackage[]{units}
\usepackage{listings}
\usepackage{multicol}
\usepackage{mathrsfs}
\usepackage{multirow}
\usepackage{tcolorbox}
\usepackage{physics}
\usepackage{subcaption,graphicx}
\usepackage{textcomp}
\usepackage{xcolor}

\title{Timing advance and Doppler shift estimation in LEO satellite networks: A recursive Bayesian study}

\author{\IEEEauthorblockN{Ashutosh Balakrishnan\textsuperscript{1}, Pierre Popineau\textsuperscript{2}, Philippe Martins\textsuperscript{1}}
\IEEEauthorblockA{\text{\textsuperscript{1}Télécom Paris, Palaiseau, France;} \textsuperscript{2}Constellation Technologies \& Operations, Paris, France \\
 balakrishnan@telecom-paris.fr, pierre.popineau@constellation.global, martins@telecom-paris.fr
}}

\begin{document}
\maketitle

\begin{abstract}
    Low earth orbit (LEO) satellite based non-terrestrial networks are a key theme of the upcoming 6G networks. These space networks are proposed to be used for high-mobility use-cases like airplanes and vehicles. 
    The initial access process between a base station (BS) and a user equipment (UE) involves timing advance (TA) value computation at the BS, requiring precise BS location information at the UE. 
    It becomes more challenging in LEO satellite networks due to the fast moving LEO satellites and large pathloss, in addition to the mobile UE. This paper aims to compute the TA and Doppler shift experienced at the UE by modeling the joint system dynamics in a LEO satellite-mobile UE network through an extended Kalman filter (EKF) based recursive Bayesian framework. 
    The framework accurately models the joint system dynamics by considering the LEO satellite acceleration. It constructs the Jacobian to linearize the inherent non-linearities present in the motion. Probabilistic insights regarding the state-update and propagation are also provided. The analytical framework factors in the limited satellite visibility at the UE and the satellite-UE geometry w.r.t. the earth center. 
    The proposed framework is also useful when the satellite and UE clocks are not in sync, with the corresponding clock drift a function of the measured time difference of arrivals. Our results showcase the efficacy and robustness of the proposed EKF framework to estimate the TA and Doppler shift, even at very high UE speeds. The work is expected to be extremely useful in realizing LEO satellite based non-terrestrial networks.
\end{abstract}



\section{Introduction}\label{SecI}

Traditionally, satellite networks have been used to aid communication in rural and far off areas devoid of terrestrial infrastructure. With the technological advancement towards deploying  low earth orbit (LEO) satellite networks (e.g., Starlink), there has been a recent rapid interest towards using these networks in urban areas \cite{Comm_surv_2023}. 

Unlike geostationary earth orbit (GEO) satellites, which are quasi-stationary to the earth reference point, LEO satellites are no longer stationary with respect to the user equipment (UE) located at the earth surface. The fast moving LEO satellites are ``visible'' only for a short time period, for which the entire network needs to be designed. In this paper, we study the problem of timing advance and Doppler shift estimation in LEO satellite networks. In the upcoming subsection, we motivate and position our work w.r.t. the current state of art.  

\subsection{Motivation}\label{SecI-A}
Timing advance (TA) is a standard initial access protocol to establish communication between the UE and the base station (BS). The initial access mechanism consists of the BS transmitting ``cell search'' signals in the downlink, followed by the UE transmitting a  handshake signal to the BS on the uplink ``random access channel (RACH)'' \cite{li2017design}. 
It may be noted that till the reception of the RACH from the UE, the BS is agnostic to the UE location. Following reception of the RACH, the BS computes a ``timing advance'' value for each UE to synchronize the uplink signals coming to it from all the UEs associated with the BS. The TA is computed based on the UE locations in the coverage area. 
The TA is then transmitted by the BS to the UE on the downlink.

The initial access procedure described above is challenging to be incorporated in LEO satellite networks. LEO satellite networks (where the LEO satellite acts as a space-BS) experience an increased path loss between LEO satellite and UE as compared to a terrestrial scenario. Further, as the LEO satellite is no longer stationary with respect to the UE, the satellite motion causes the initial access signals to experience Doppler shift. The problem becomes more challenging if the ground UE is also mobile. 

\begin{figure*}[t]
\centering
\subfloat[][]{
\includegraphics[width=0.33\textwidth]{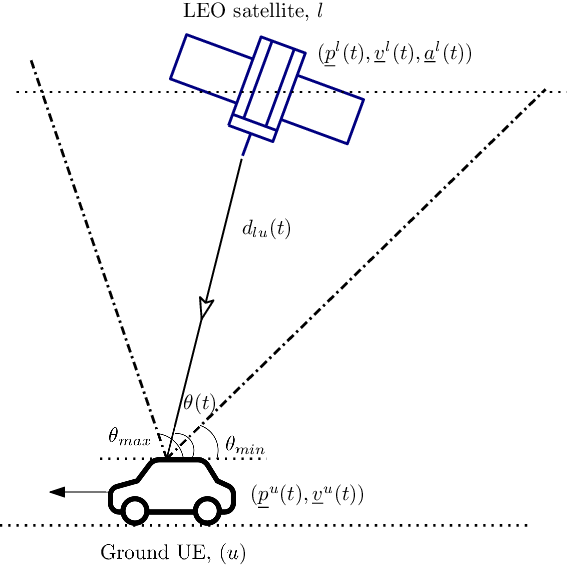}}
\subfloat[][]{
\includegraphics[width = 0.42\textwidth]{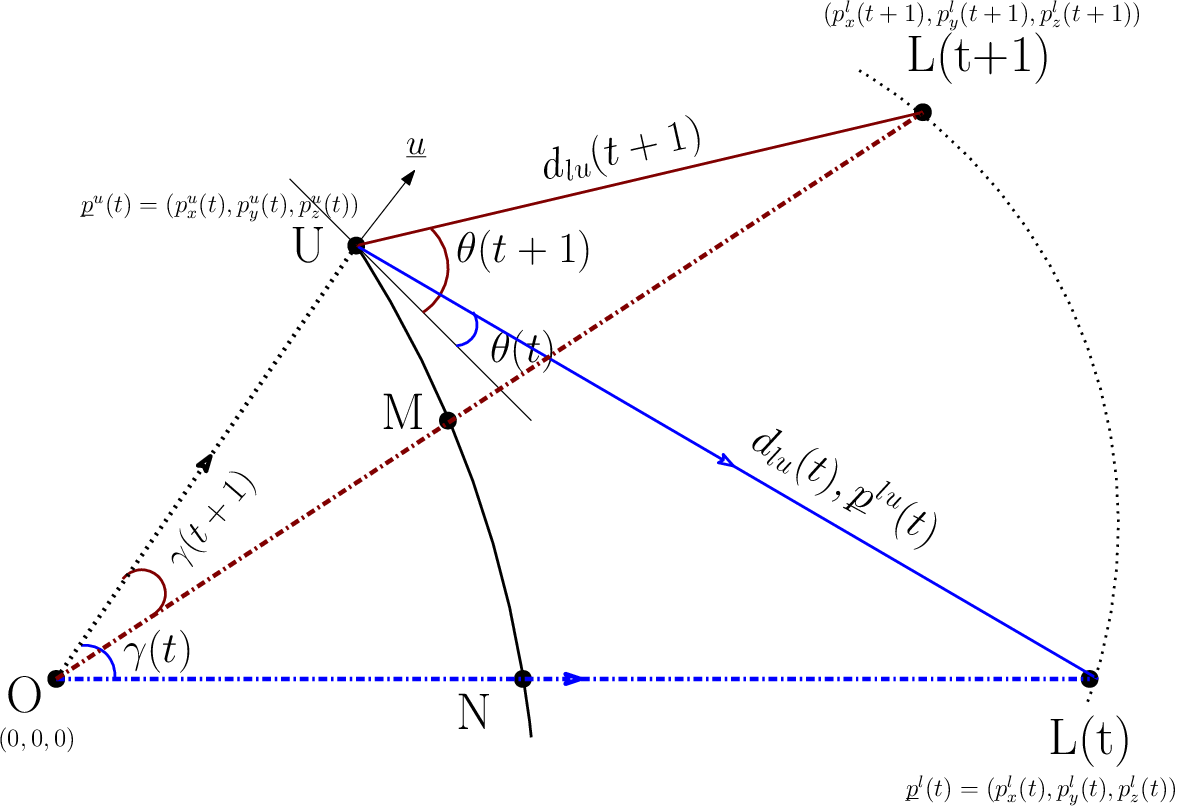}}
\vspace{-2mm}
\caption{Illustration of (a) LEO satellite-terrestrial UE based communication network and (b) geometry of the considered system depicting a UE $u$ (location $\mathbf{U}$), LEO satellite $l$ (location $\mathbf{L}$) with respect to the earth center $\mathbf{O} \in \mathbb{R}^3$ having coordinates $(0,0,0)$.}\label{fig:1}
\end{figure*}

\subsection{Related works}
Accurate estimation of the TA requires accurate estimate of the LEO satellite location as well as the UE location. Recently, the authors in \cite{Access_2019} presents a signal level analysis to capture the timing and frequency offset.  The frameworks in \cite{GLOBECOM_2020}, \cite{TVT_2021} present a location based TA estimation algorithm.  The authors in \cite{TVT_2024} consider TA estimation assuming that either of UE/LEO location as well known. 
The authors in \cite{Aerospace_transaction}  present an algorithmic extended Kalman filter (EKF) based ephemeris tracking framework. The states tracked include only position of satellite. The satellite visibility is not considered with respect to communication. The UE is also assumed static w.r.t. satellite. The probabilistic analysis concerning the filtering and satellite ephemeris is ignored.

The frameworks in \cite{Access_2019, GLOBECOM_2020, TVT_2021, TVT_2024} have ignored the motion of satellite, and its effect on the TA computation. Furthermore, the frameworks assume the UE to be static, and hence ignore the effect of UE mobility on TA computation. The framework in \cite{Access_2019} also assumes accurate knowledge of UE as well as satellite position, which may not be true in urban scenarios. The framework in \cite{TVT_2021} has ignored the limited LEO sat. visibility as well
Motivated by the above problems, in this work we propose an EKF based recursive Bayesian framework to estimate the LEO satellite position and velocity coordinates, in addition to capturing the UE dynamics. Below we summarize the key contributions through this work.

\subsection{Contributions}\label{SecI-B}
\begin{enumerate}
        \item In this paper we aim to capture and study the effects of joint dynamics of the LEO satellite-ground UE through an EKF based recursive Bayesian framework. 
        \item First, we begin by  studying the motion model of the joint LEO Satellite-UE dynamics and construct the Jacobian to linearize the inherent non-linearities present in the motion model. The system stability is analyzed through the algebraic Riccatti equations in the continuous time domain. Probabilistic insights regarding the state-update and propagation are provided.  
        \item Factoring in the limited satellite visibility of the UE, the satellite-UE geometry with respect to the earth, and the measurements obtained through the proposed EKF framework, the TA value is computed. The Doppler shift is computed through the the time difference of arrivals computed at various positions of the LEO satellite in the visibility period. It is shown when the clock of the LEO satellite and UE are not in sync, the clock drift is a function of the time difference of arrivals.  
        \item Our results illustrates the robustness and efficacy of the proposed EKF based Bayesian framework, by achieving mean percentage error of $[1.8166\% \ 0.5595\% \ 0.7725\%]$ in estimating the LEO satellite position coordinates even at high speeds of the UE. The work also demonstrates the accurate computation of the TA \& Doppler shift and the effect of UE mobility on key communication performance parameters.  
    \end{enumerate}

\subsection{Organization}\label{SecI-C}
Section \ref{SecII} presents the LEO sat.-- UE based system model. Section \ref{SecIII} outlines the Bayesian approach to solve the problem. Section \ref{SecIV} studies the TA and Doppler shift computations. Section \ref{SecV} discusses the results and observations followed by Section \ref{SecVI} which concludes the paper. 

\section{System Model}\label{SecII}

The system model comprises of a single LEO satellite $l$, having position, velocity, and acceleration $\{\underbar{$p$}^l, \underbar{$v$}^l, \underbar{$a$}^l\} \in \mathbb{R}^3$, deployed on an orbit with known ephemeris. The LEO satellite architecture is a regenerative architecture with the satellite acting as a space-BS. The system also considers a single UE $u$ on the earth's surface having position and velocity $\{\underbar{$p$}^u, \underbar{$v$}^u\} \in \mathbb{R}^3$, as shown in Fig. \ref{fig:1}(a). It may be noted that the position, velocity, and acceleration are w.r.t. the earth center $\mathbf{O}$ denoting the origin in $\mathbb{R}^3$. The earth-centered angle, $\gamma {=} \measuredangle LOU$, refers to the angle subtended between the  satellite-earth center-UE as shown in Fig. \ref{fig:1}(b). The elevation angle $\theta$ corresponds to the angle between the satellite  and the tangent drawn at earth surface. The slant range $d_{lu}(t)$ refers to the radial distance between the user $u$ and satellite $l$. As the satellite moves in space with time (say to $l(t+1)$), the earth-centered angle, elevation angle, and the slant range change correspondingly. The maximum and minimum elevation angle corresponding to the LEO satellite visibility from the UE is denoted as $\theta_{max}$ and $\theta_{min}$.
 
Let $\underbar{$p$}^u {=} (\phi^u, \lambda^u, h^u)$ represent the location of UE $u$ in geodetic coordinates, by capturing the UE latitude, longitude, and height, respectively.  
We first transform the user geodetic coordinates to geo-centric coordinates $(p_x^u, p_y^u, p_z^u)$ (i.e., with respect to the earth center) using the below equations \cite{curtis2019orbital}.
\begin{equation}
    \begin{aligned}
        x^u &= (N+h)\cos \phi \cos \lambda \\
        y^u &= (N+h)\cos \phi \sin \lambda \\
        z^u &= \frac{b^2 (N+h)}{a^2} \sin \phi.
    \end{aligned}
\end{equation}
Here, $a$ and $b$ denote the equatorial (semi-major-axis) and polar radius (semi-minor axis) of the earth, $N$ refers to the prime-vertical radius of curvature given by
\begin{equation}
    N = a^2 \slash (\sqrt{a^2 \cos^2 \phi + b^2 \sin^2 \phi}). 
\end{equation}
In this work, for analytical simplicity, we consider the earth to be spherical (and not ellipsoidal) having $a{=}b{=}R_e$, with $R_e$ denoting the earth's radius. The user is assumed to move with a velocity $\underbar{$v$}^u {=} (v^u_x, v^u_y, v^u_z)$, w.r.t. the earth center. 
The general discrete-time joint system dynamics of the LEO satellite--UE network is given as,
\begin{equation}\label{gen_sys}
    \underbar{$x$}_{t+1} = \mathcal{F}(\underbar{$x$}_t)  + \underbar{$\nu$}_t, 
\end{equation}
\begin{equation}\label{gen_meas}
    \underbar{$y$}_t = h(\underbar{$x$}_t) + \underbar{e}_t.
\end{equation}
Here, \eqref{gen_sys} and  \eqref{gen_meas} correspond to the motion and  the measurement model of the system, respectively.  $\underbar{$x$}_t$ refers to the state vector to be estimated assuming the initial state $\underbar{$x$}_0 {\sim} f_{\underbar{$x$}_0}()$ to be known. 
$\underbar{$\nu$}_t {\sim} f_\nu()$ and $e_t {\sim} f_{e}()$ denote the additive noise in the motion and measurement processes. $\underbar{$y$}_t$ refers to the observable measurement, and $h()$ is a deterministic function of the state $\underbar{$x$}_t$.   

The distributions $f_\nu(), f_{e}()$ are assumed to be Gaussian and independent to each other. The system also follows Markovian property, i.e., $\underbar{$x$}_{t+1}$ depends only on $\underbar{$x$}_t$. These properties make such systems easily realizable and not computationally intensive. The aim of the framework is to observe the measurements at time $t, Y^t {=} \{y_0, y_1, \ldots, \underbar{$y$}_t\}$ and make an intelligent estimate of $\underbar{$x$}_{t+1}$.  In the upcoming subsections, we present the motion and measurement models in detail.

\subsection{Motion model}\label{SecII-A}

We consider the LEO satellite to be subjected to a constant acceleration based time varying motion model. For a LEO satellite $l$ deployed in earth's orbit with coordinates $\underbar{$p$}^l {=} (p_x^l, p_y^l, p_z^l)$, the primary acceleration due to earth's gravitational potential \cite{vinti1998orbital} is given as
\begin{equation}\label{grav_pot}
    \underbar{a}^l = - \mu \ \underbar{$p$}^l \slash \norm{\underbar{$p$}^l}^3.
\end{equation}
Here, $\underbar{a}^l {=} (a^l_x, a^l_y, a^l_z)$, $\mu$ refers to the earth's gravitational constant, and $\norm{\underbar{$p$}^l} {=} \sqrt{(p_x^l)^2 + (p_y^l)^2 +(p_z^l)^2 }$ refers to the distance of the satellite from earth center. 

Let the state space correspond to the position and velocity of the LEO satellite $l$ given as $\underbar{$x$}^l {=} [\underbar{$p$}^l \ \underbar{$v$}^l]^T {=} [p_x^l, p_y^l, p_z^l, v^l_x, v^l_y, v^l_z]^T$. It may be noted that $(\cdot)^T$ refers to the transpose operator. The motion model corresponding to the states are,
\begin{equation}
    \underbrace{\begin{bmatrix}
        p_i^l(t+1)  \\
        v_i^l(t+1) 
        \end{bmatrix}}_{\underbar{$x$}_{t+1}}
        =
        \underbrace{\begin{bmatrix}
            p_i^l(t) + v^l_i \Delta t \\
            v_i^l(t) + a_i^l \Delta t 
        \end{bmatrix}}_{\mathcal{F}(\underbar{$x$}_t)}
        +
        \underbar{$\nu$}_t, \ \forall \ i \in \{x, y, z\}.
\end{equation} 

Since the LEO sat. motion in \eqref{grav_pot} introduces non-linearity with respect to the state variable, we linearize the motion model by constructing the Jacobian, $\underbar{$F$} (t)  {=} \partial \mathcal{F}\slash \partial \underbar{$x$}^l$ as,
\begin{equation}\label{Jacob1}
    \underbar{$F$} (t)  = 
    \begin{bmatrix}
    \frac{\partial  p_x^l}{\partial  \underbar{$x$}^l} & \frac{\partial  p_y^l}{\partial  \underbar{$x$}^l} & \frac{\partial  p_z^l}{\partial  \underbar{$x$}^l} & \frac{\partial  v_x^l}{\partial  \underbar{$x$}^l} & \frac{\partial  v_y^l}{\partial  \underbar{$x$}^l} & \frac{\partial  v_z^l}{\partial  \underbar{$x$}^l}
    \end{bmatrix}^T_{6\times6}.
\end{equation}
The partial derivatives of the position coordinates with respect to the state variables result as follows: $(\partial p_i^l \slash \partial p_i^l) {=} 1$ and $(\partial p_i^l \slash \partial v_i^l) {=} \Delta t \ \forall \ i \in \{x, y, z\}$. To solve for the partial derivatives of the velocity coordinates with respect to the state variables, we illustrate the computation for one case below
\begin{equation}
\begin{aligned}
    \partial v_x^l \slash \partial p_x^l &= -\mu \frac{\partial}{\partial  p_x^l } \left( \frac{p_x^l}{((p_x^l)^2 + (p_y^l)^2 +(p_z^l)^2)^{3/2}} \right) \\
    &= \frac{-\mu }{\norm{\underbar{$p$}^l}^5}( \norm{\underbar{$p$}^l}^2 - 3 (p_x^l)^2).
\end{aligned}
\end{equation}
Hence, 
\begin{equation}
    \begin{aligned}
        \partial v_i^l \slash \partial p_i^l &= \frac{-\mu }{\norm{\underbar{$p$}^l}^5}( \norm{\underbar{$p$}^l}^2 - 3 (p_i^l)^2) \Delta t \\
        \partial v_i^l \slash \partial p_j^l &= \frac{\mu }{\norm{\underbar{$p$}^l}^5}(3ij) \Delta t \ \forall i\neq j
    \end{aligned}
\end{equation}
Furthermore, $\partial v_i^l \slash \partial v_i^l {=} 1$ and $\partial v_i^l \slash \partial v_j^l {=} 0$. The Jacobian $\underbar{$F$} (t) $ in \eqref{Jacob1} can be written as, 
\begin{equation}
    \underbar{$F$} (t)  = 
    \begin{bmatrix}
    I_{3\times 3} & \Delta t I_{3\times 3} \\
    \mathbb{A} \Delta t_{3\times 3}  & I_{3 \times 3}
    \end{bmatrix}
\end{equation}
where, 
\begin{equation}
\resizebox{\hsize}{!}{$
\mathbb{A} = \frac{-\mu }{\norm{\underbar{$p$}^l}^5}
    \begin{bmatrix}
        (\norm{\underbar{$p$}^l}^2 - 3 (p_x^l)^2) & -3p_x^l p_y^l & -3p_x^l p_z^l  \\
        -3p_y^l p_x^l & (\norm{\underbar{$p$}^l}^2 - 3 (p_y^l)^2) & -3p_y^l p_z^l  \\
        -3p_z^l p_x^l & -3 p_z^l p_y^l & (\norm{\underbar{$p$}^l}^2 - 3 (p_z^l)^2)
    \end{bmatrix}. $}
\end{equation}

To capture the motion of the UE as well as the LEO satellite jointly, the state space is modified as, $\underbar{$x$}^{lu} {=} [\underbar{$x$}^l \ \underbar{$p$}^u \ \underbar{$v$}^u]^T$. The corresponding motion model can be written as,
\begin{equation}
    \underbrace{\begin{bmatrix}
        p_i^l(t+1)  \\
        v_i^l(t+1)  \\
        p_i^u(t+1)  \\
        v_i^u(t+1)
        \end{bmatrix}}_{\underbar{$x$}_{t+1}}
        =
        \underbrace{\begin{bmatrix}
            p_i^l(t) + v^l_i \Delta t \\
            v_i^l(t) + a_i^l \Delta t \\
            p_i^u(t) + v^u_i \Delta t \\
            v_i^u(t)
        \end{bmatrix}}_{\mathcal{F}(\underbar{$x$}_t)}
        +
        \underbar{$\nu$}_t, \ \forall \ i \in \{x, y, z\}.
\end{equation} 
It may be noted that the UE is assumed to move with constant acceleration and unlike for the satellite, the UE acceleration does not get affected by the UE position on the earth surface. The Jacobian for the joint UE-satellite model can be constructed as in (10) and is given as,
\begin{equation}\label{jacob2}
\underbar{$F$} (t)  = 
    \begin{bmatrix}
        I_{3\times 3} & \Delta t I_{3\times 3} & 0_{3 \times 3} & 0_{3\times 3} \\
        \mathbb{A} \Delta t_{3\times 3}  & I_{3 \times 3} & 0_{3 \times 3} & 0_{3\times 3} \\
        0_{3 \times 3} & 0_{3\times 3} & I_{3\times 3} & \Delta t I_{3\times 3} \\
        0_{3 \times 3} & 0_{3 \times 3} & 0_{3 \times 3} & I_{3\times 3}
    \end{bmatrix}.
\end{equation}

\subsection{Measurement model}\label{SecII-B}
This subsection outlines the measurements $\underbar{$y$}_t$, which will be used to estimate the state vector $\underbar{$x$}_t$. The range and elevation angle between the UE and the LEO satellite can be a very useful measurements.
\begin{enumerate}
        \item Range, $d_{lu}(\underbar{$p$}^l (t), \underbar{$p$}^u (t)) = \norm{\underbar{$p$}^l{-}\underbar{$p$}^u}$ is the euclidean distance between the UE and LEO satellite.
        
        \item Elevation angle, $\theta(\underbar{$p$}^l(t), \underbar{$p$}^u(t))$ is  defined as the satellite's visibility angle with respect to the user's local horizon. Let $\underbar{$p$}^{lu} (t) {=} (\underbar{$p$}^l(t) - \underbar{$p$}^u(t))$. Furthermore, the unit vector along the user's local zenith direction is given as, $\underbar{$u$} {=} \underbar{$p$}^u\slash \norm{p^u}$. Then, 
\begin{equation}
    \theta(t) = \sin^{-1}\left( \frac{\underbar{$p$}^{lu}(t) \cdot\underbar{$u$}(t)}{\norm{\underbar{$p$}^{lu}(t)}} \right).
\end{equation}
where $(\cdot)$ denotes the dot product operator.
    \end{enumerate}
In addition to the range and elevation angle, the Doppler shift and the clock drift being functions of the state-space variables, can also be measured accurately (discussed later in Section \ref{SecIV}). In the upcoming section, we present the proposed recursive Bayesian framework to estimate the state-variables.


\section{Recursive Bayesian Framework}\label{SecIII}
In this section we first outline the estimation approach employed for linear time varying systems termed as Kalman filtering (KF), followed by the  linearized version of KF used for non-linear systems, termed as extended KF. Finally, we present the probabilistic insights gained through the study.  

\subsection{Optimal Kalman filtering}\label{SecIII-A}
Kalman filtering (KF) first involves computing the estimate $\underbar{$\hat{x}$}_{t-1}$ using the measurement/ observation samples up to time $(t-1)$, i.e.,  $Y^{t-1} {=} \{\underbar{$y$}_0,\underbar{$y$}_1, \ldots, \underbar{$y$}_{t-1}\}$. It may be noted that the process and measurement noises are assumed to be Gaussian distributed, i.e., $\underbar{$\nu$}_t {\sim} N(0, \underbar{$Q$}_t)$ and $\underbar{$e$}_t {\sim} N(0, \underbar{$R$}_t)$. 
Both these noises are independent and uncorrelated with each other, i.e., $\mathbb{E}(\underbar{$\nu$}_{j} \ \underbar{$\nu$}_{k}) {=} 0, \mathbb{E}(\underbar{$e$}_{j} \ \underbar{$e$}_{k}) {=} 0, \mathbb{E}(\underbar{$\nu$}_{j} \ \underbar{$e$}_{k}) {=} 0 \ \forall \ j, k \in t; j\neq k$.  With initial conditions, $\underbar{$x$}_0 \sim N(0, \underbar{$P$}_{1|0})$, the Kalman gain $\underbar{$K$}_t$ corresponding to the minimum mean squared error (MMSE) problem formulation is given by, 
\begin{equation}\label{mmse}
    \underbar{$K$}_t = \arg \min_{\underbar{$K$}_t} \mathbb{E}[|\hat{\underbar{$x$}} - \underbar{$x$}|^2].
\end{equation}

For a linear time varying system, $\mathcal{F}(\underbar{$x$}_t)$ in \eqref{gen_sys} and $h(\underbar{$x$}_t)$ in \eqref{gen_meas} can be represented in matrix form as $\underbar{$A$}_t \underbar{$x$}_t $ and $\underbar{$H$}_t \underbar{$x$}_t $.  The KF equations for state propagation and update corresponding to solving the MMSE problem are given as,
%
    \begin{align}
        \hat{\underbar{$x$}}_{t|t-1} &= \underbar{$A$}_{t-1} \ \hat{\underbar{$x$}}_{t-1}, \ \text{where} \  \hat{\underbar{$x$}}_{t-1} = \mathbb{E}[\underbar{$x$}_{t-1}] \label{KF1}\\
        \hat{\underbar{$y$}}_t &= \underbar{$H$}_t \  \hat{\underbar{$x$}}_{t|t-1} \label{KF2}\\
        \hat{\underbar{$x$}}_{t} &= \hat{\underbar{$x$}}_{t|t-1} + \underbar{$K$}_t[\underbar{$y$}_t - \hat{\underbar{$y$}}_t] \label{KF3}\\
        \text{and} \ \underbar{$K$}_t &= \underbar{$P$}_{t|t-1} \underbar{$H$}_t^T(\underbar{$H$}_t \ \underbar{$P$}_{t|t-1} 
 \ \underbar{$H$}_t^T + \underbar{$R$}_t )^{-1}. \label{KF4}
    \end{align}
%
The covariance propagation and update equations corresponding to the MMSE problem are given as,
    \begin{align}
        \underbar{$P$}_{t+1|t} &= \underbar{$A$}_t \underbar{$P$}_t  \underbar{$A$}_t^T + \underbar{$Q$}_t  \label{KF5}\\
        \underbar{$P$}_t  &= (I - \underbar{$K$}_t \ \underbar{$H$}_t) \ \underbar{$P$}_{t|t-1}. \label{KF6}
    \end{align}
Here, $\underbar{$P$}_{t|t-1} = \mathbb{E}[(\hat{\underbar{$x$}}_{t|t-1} - \underbar{$x$}_t)(\hat{\underbar{$x$}}_{t|t-1} - \underbar{$x$}_t)^T]$ and $\underbar{$P$}_t  = \mathbb{E}[(\hat{\underbar{$x$}}_{t} - \underbar{$x$}_t)(\hat{\underbar{$x$}}_{t} - \underbar{$x$}_t)^T]$. The probabilistic insights of these expressions will be explained shortly in Section \ref{SecIII-C}.

\subsection{Extended Kalman filtering}\label{SecIII-B}
Considering the non-linearities present in the system dynamics due to the acceleration experienced by the satellite, we linearize the system dynamics through the Jacobian as illustrated in \eqref{Jacob1} and \eqref{jacob2}, and this linearized version of KF is termed as ``extended Kalman filter, EKF''. The EKF works similar to the KF, but for a non-linear time varying system wherein $\mathcal{F}(\underbar{$x$}_t)$ in \eqref{gen_sys} can be represented in matrix form as $\underbar{$F$}_t \underbar{$x$}_t $. Here, $\underbar{$F$}_t$ represents the Jacobian constructed in \eqref{jacob2}. 

Hence, the equations \eqref{KF1}-\eqref{KF6} are similar for the EKF with a subtle modification of $\underbar{$A$}_t \rightarrow \underbar{$F$}_t$, wherever applicable. It may be noted that since optimality of the MMSE problem in \eqref{mmse} is not guaranteed through the linearization based EKF, hence we check the stability of the system at steady state through the Algebraic Riccatti equations \cite{anderson2005optimal} given as 
\begin{equation}
    \underbar{$P$}_{\infty} = \frac{\underbar{$R$}}{\underbar{$H$}^2}\left(\underbar{$F$} + \sqrt{\underbar{$F$}^2 + \underbar{$H$}^2 \underbar{$Q$} \slash \underbar{$R$}} \right).
\end{equation}
In the upcoming section, we will provide insights on the probabilistic aspects of the presented analysis.

\subsection{Probabilistic analysis}\label{SecIII-C}
The probabilistic aspects of the above filtering can be summarized in four key steps as provided below. 
\begin{enumerate}
        \item Initial conditions, $f(\underbar{$x$}_0), \underbar{$P$}_{1|0}$
        \item a-priori belief propagation, 
        \begin{equation}\label{aprioi}
            f_{t|t-1}(\underbar{$x$}_t) {=} {\int_{-\infty}^{\infty}} f_v(\underbar{$x$}_t {-} \underbar{$F$}_{t-1} \underbar{$x$}_{t-1}) f_{t-1}(\underbar{$x$}_{t-1}) d\underbar{$x$}_{t-1}
        \end{equation}
        \item Obtain $\underbar{$y$}_t$, compute a-posteriori belief: 
        \begin{equation}\label{aposteriori}
            f_t(\underbar{$x$}_t) = f_e(\underbar{$y$}_t - h(\underbar{$x$}_t)) f_{t|t-1}(\underbar{$x$}_t)
        \end{equation}
        \item Return $f_t(\underbar{$x$}_t)$, set $t=(t+1)$ and repeat.
    \end{enumerate}
    With knowledge of the initial conditions $f(\underbar{$x$}_0), \underbar{$P$}_{1|0}$, the framework first involves computation of $\hat{\underbar{$x$}}_{t-1}$ having a distribution $f_{t-1}(\underbar{$x$}_{t-1})$ (refer \eqref{KF1}). This statistic is used to recursively compute  the a-priori belief distribution $f_{t|t-1}(\underbar{$x$}_t)$ associated with $\hat{\underbar{$x$}}_{t|t-1}$ (refer \eqref{KF1} and \eqref{aprioi}). This step is followed by the computation of the predicted output $\hat{\underbar{$y$}}_t$ in \eqref{KF2}. After assimilating a new measurement $\underbar{$y$}_t$, the framework then uses the a-priori belief to compute the a-posteriori belief (as in \eqref{aposteriori}). In the filtering process, this step involves the correction of  $\hat{\underbar{$x$}}_{t}$ (a-posteriori estimate) using $\hat{\underbar{$x$}}_{t|t-1}$ and $\hat{\underbar{$y$}}_t$, as shown in \eqref{KF3}. The Kalman gain computed in this process is computed through \eqref{KF4}.

We use the EKF framework presented to estimate key communication parameters as show in the coming section. 

\section{TA and Doppler Shift Analysis }\label{SecIV}
In this subsection we delineate the computation process involved behind estimating the timing advance value and Doppler shift experienced by a ground UE relative to a LEO satellite, w.r.t. the earth center.

\subsection{Timing advance}\label{SecIV-A}
 
To compute the TA value, we first understand the earth-centered angle. With respect to Fig. \ref{fig:1}(b), we first define the earth-centered angle $\gamma$ and elevation angle $\theta$, respectively. It may be noted that $\gamma$ refers to the central angle subtended between the UE and satellite through the earth center. With the coordinates of the UE and satellite at time $t$ denoted by $\underbar{$p$}^u(t) {=} (x^u_t, y^u_t, z^u_t)$ and $\underbar{$p$}^l(t) {=} (x^l_t, y^l_t, z^l_t)$,
\begin{equation}
    \gamma(t) = \cos^{-1}{\frac{\underbar{$p$}^u(t) \cdot \underbar{$p$}^l(t)}{\norm{\underbar{$p$}^u(t)} \norm{\underbar{$p$}^l(t)}}}.
\end{equation}
This calculation directly uses the dot product between two vectors $\underbar{$p$}^u(t), \underbar{$p$}^l(t)$ w.r.t. the earth center.  Here, $\norm{k}$ represents the $l_2$ norm (or magnitude) of the vector and $(\cdot)$ represents the dot product operator. 
For the triangle $LOU$, the slant range $d_{lu}(t)$ (as a function of the earth-centered angle) is determined by applying cosine law as, 
    $d_{lu}(t) {=} \sqrt{R_e^2 {+} \norm{\underbar{$p$}^l_t}^2 {-} 2R_e \norm{\underbar{$p$}^l_t} \cos{\gamma_t} }.$ 
The TA being the total round trip time between UE and the LEO satellite, as a function of the earth-centered angle and satellite position is 
\begin{equation}
    TA(t) = 2. d_{lu}(t) \slash c. 
\end{equation}
The error in timing advance is given as $(d_{lu}-\hat{d}_{lu})^2$, where $d_{lu}$ and $\hat{d}_{lu}$ represent the true and estimated slant range. 

\subsection{Doppler shift and Clock drift}\label{SecIV-B}
The Doppler shift experienced by the UE is given as, $\Delta f(t) {=} f_T \frac{\underbar{$v$}^l(t) - \underbar{$v$}^u(t)}{c}$, where $f_T$ is transmitted signal frequency and $(\underbar{$v$}^l(t) {-} \underbar{$v$}^u(t))$ is relative velocity between satellite and UE.
Let, $\Delta \tau_1 {=} (t_{12} {-} t_{11}) = (d_{lu}(t+1){-}d_{lu}(t)) \slash c)$ represent the time difference of arrivals (TDoA) between the signals coming from the LEO sat. to the UE at two time samples $(t+1)$ and $t$. It is currently assumed the clocks of LEO satellite and UE are synchronized. 
Taking time derivative, 
\begin{equation}
\begin{aligned}
    \frac{ \dd \Delta \tau_1}{\dd t} &= \frac{\dd \slash \dd t (d_{lu}(t+1)) - \dd \slash \dd t (d_{lu}(t))}{c} \\
    &= \frac{(\underbar{$v$}^l(t+1) - \underbar{$v$}^u(t+1)) - (\underbar{$v$}^l(t) - \underbar{$v$}^u(t))}{c} \\
    &= (\Delta f(t+1)- \Delta f(t)) \slash f_T.
    \end{aligned}
\end{equation}
Hence, accurate estimation of the LEO satellite--UE position and velocity through the proposed framework is very useful to gather engineering insights about the frequency difference of arrivals and thereby the rate of change of Doppler shift. 

Let the LEO satellite and UE clocks not be in sync, thereby indicating the presence of a clock drift. Let $t_{11} \epsilon_1$ denote the initial clock offset and $\epsilon_2$ denote the clock drift at time $t_{12}$. The modified time of arrival of the signals in this non-synchronized scenario can be given as,
    \begin{equation}
    \begin{aligned}
        \hat{t}_{12} {=} t_{12}(1 + \epsilon_2), \      \hat{t}_{11} {=} t_{11}(1 + \epsilon_1)
    \end{aligned}
    \end{equation}
    Hence, erroneous TDoA is given as,
    \begin{equation}
        \begin{aligned}
            \Delta \hat{\tau}_1 {=} (\hat{t}_{12} {-} \hat{t}_{11}) 
            {=} (t_{12} {-} t_{11}) {+} t_{12}\epsilon_2 {-} t_{11} \epsilon_1.
        \end{aligned}
    \end{equation}
Taking difference of the erroneous TDoA and the ideal TDoA, we get the corresponding clock drift between the two time intervals which is given as  
\begin{equation}
 \Delta t_{\text{clock drift}} {=} \Delta \hat{\tau}_1 - \Delta \tau_1 = (t_{12}\epsilon_2 - t_{11} \epsilon_1 )  {=}  t \alpha + \beta
\end{equation}
   Hence, it is inferred that $\Delta t_{\text{clock drift}}$ is of the form $\alpha t {+} \beta$, with the corresponding rate of change of clock drift being, $\dd \slash \dd t (\Delta t_{\text{clock drift}}) {=} \alpha$. Thus, the proposed EKF framework is extremely potent to compute the TA, Doppler shift and even the clock drift. Accurate estimation of these parameters are key to realizing a robust LEO satellite aided network. We discuss the results and inferences in the upcoming section.

\section{Results}\label{SecV}

\begin{figure*}[t]
\centering
\subfloat[][]{
\includegraphics[width = 0.33\textwidth]{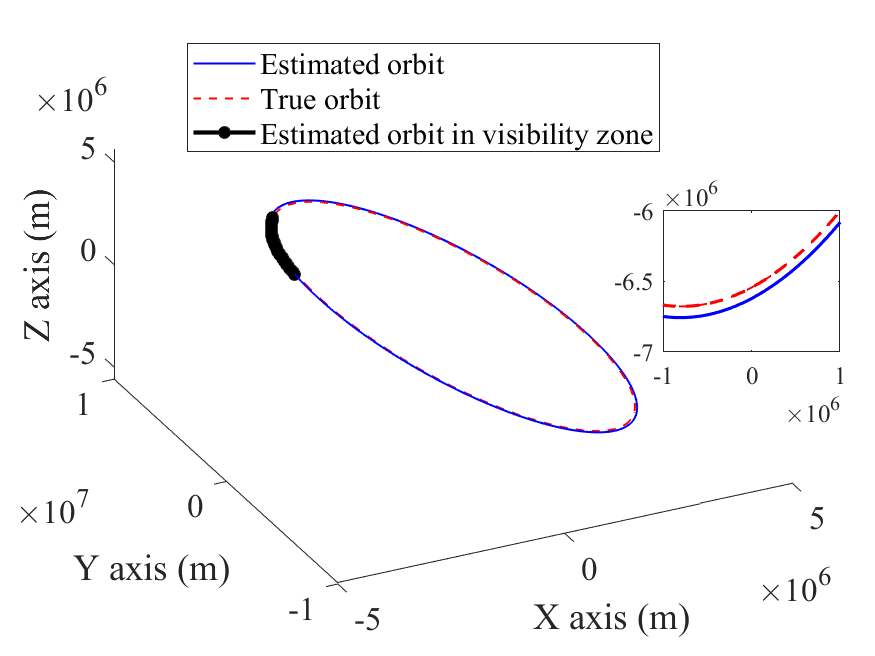}}
\subfloat[][]{
\includegraphics[width = 0.33\textwidth]{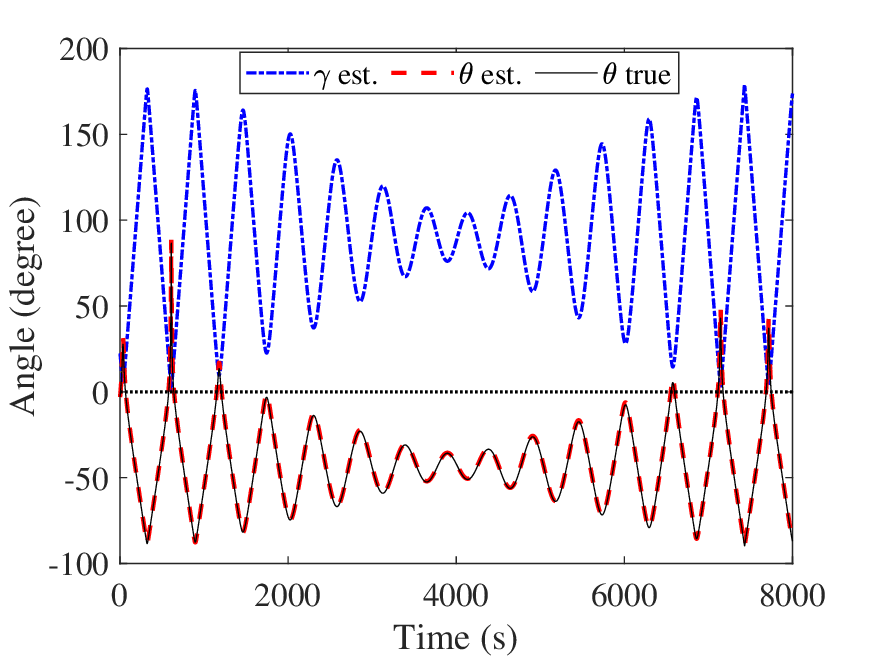}}
\subfloat[][]{
\includegraphics[width = 0.33\textwidth]{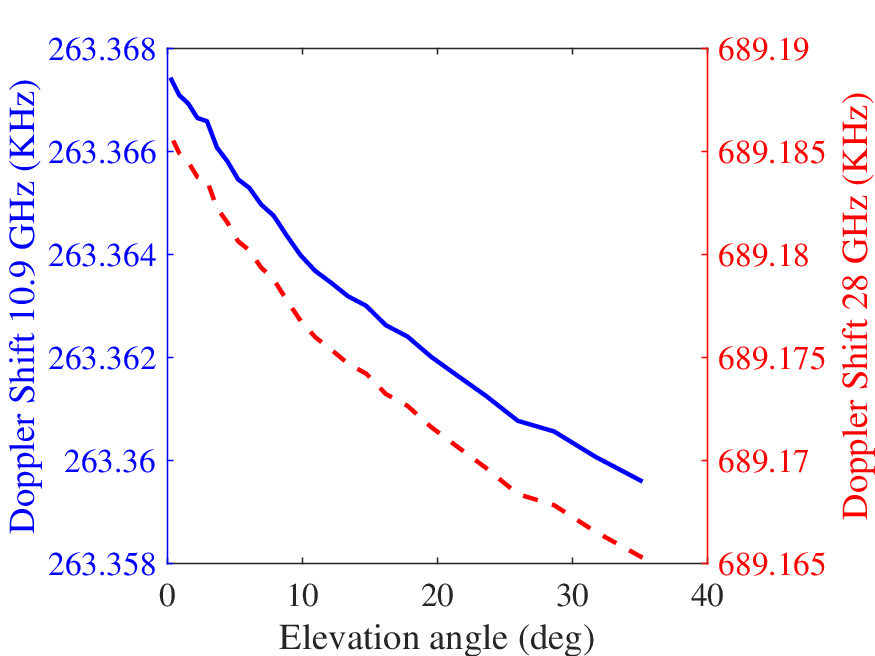}}
\caption{Illustrating (a) estimated and measured coordinates of the LEO satellite in space. The orbit portion in which the satellite is visible from Paris (the city considered for this experiment) is also highlighted.  (b) variation of estimated earth-centered angle $\gamma$ and elevation angle (true \& estimated) $\theta$ with time. It may be noted that the satellite is visible for $\theta \geq 0$. (c) Doppler shift variation with elevation angle at frequencies $10.9 \mbox{GHz}$ and $28 \mbox{GHz}$. }\label{fig:2}
\end{figure*}

\begin{figure}[t]
\centering
\subfloat[][]{
\includegraphics[width = 0.25\textwidth]{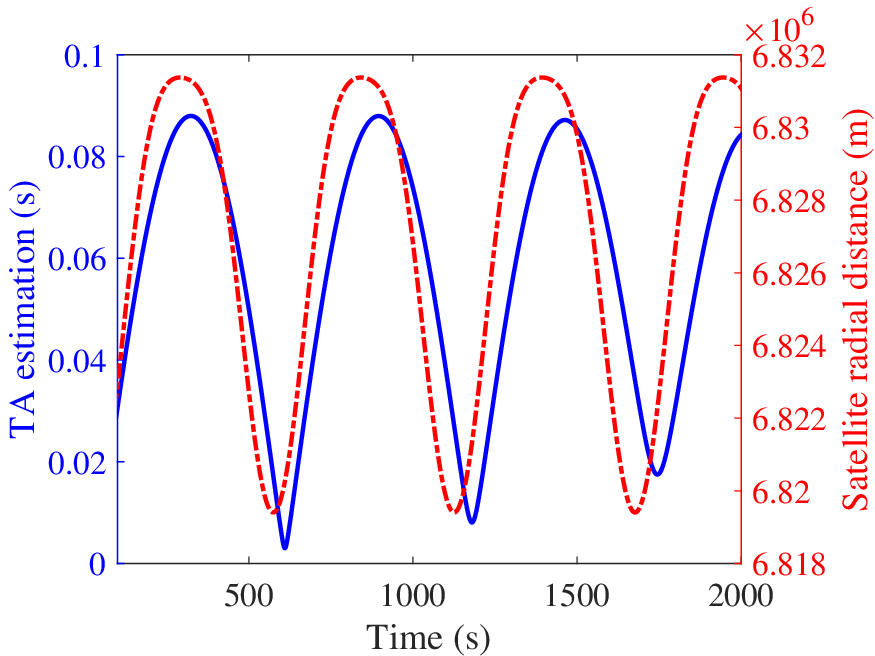}}
\subfloat[][]{
\includegraphics[width = 0.25\textwidth]{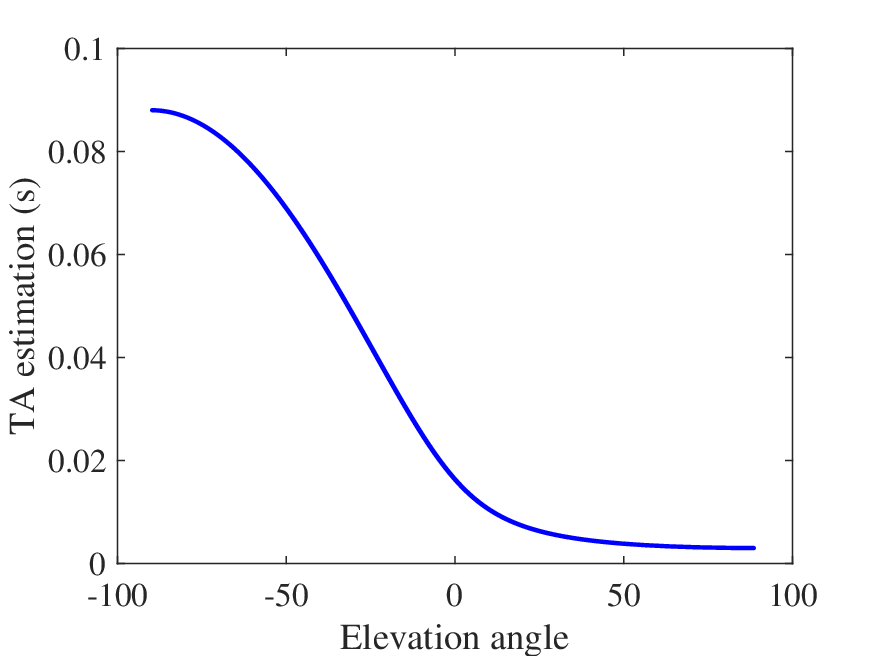}}
\caption{(a) Illustrating variation of TA and radial distance of satellite from earth center. (b) Illustrating variation of TA with elevation angle.}\label{fig:3}
\end{figure}

In this section we illustrate the key results through experimental simulations. The framework has been simulated for a UE located at INRIA Paris, France, with latitude $\phi^u {=} 48.8323 \deg$ and longitude $\lambda^u = 2.3364 \deg$. The UE is assumed to move in a linear motion with a fixed velocity. It may be noted that while the simulations have been performed with constant velocities, the analytical framework can handle time varying as well as non-linear UE motions as well. 

The simulation values are, $i{=}55\deg$, $\norm{\underline{v}^u} {=} 306.77\mbox{mps},$ $f_T{=}10.7 \mbox{GHz}$, $R_e{=}6371 \mbox{Km}$, $c{=}3e8\mbox{mps}$, $\Delta t{=}10\mbox{ms}$, $\sigma_{iQ}^2{=}10^{-4}\mbox{Km/s}^2$,  $\sigma_{iR}^2{=}0.1 \mbox{Km}$.
We have obtained $10000$ samples of position, velocity, and acceleration data, of a LEO satellite that Constellation Technologies \& Operations, France aims at deploying on a $55$ deg shell at $375\mbox{Km}$ altitude. 

First we illustrate the efficacy of the proposed Bayesian framework through Fig. \ref{fig:2}(a). Fig. \ref{fig:2}(a) depicts the estimated and true coordinates of the LEO satellite on the X, Y and Z axes. It can be seen the estimated position coordinates closely follow the measured coordinates. The mean percentage error in estimating the position coordinates comes out to be $[1.8166\% \ 0.5595\% \ 0.7725\%]$ in the respective coordinates. The orbit portion which is visible to UE is also shown.

In Fig. \ref{fig:2}(b) we illustrate the variation of the estimated  earth-centered angle $\gamma$ and elevation angle $\theta$ (true as well as estimated) for a mobile UE, moving with average speed $\norm{\underbar{$v$}^u} {=} 306.77 \mbox{mps}$ (the appx. speed of an airplane). It is observed that for a visibility period, the earth-centered angle keeps increasing (and the corresponding $\theta$ decreases). Following the min. elevation point, $\theta$ starts increasing  till the maximum elevation angle $\theta_{max}$, and the cycle repeats. The trend of variation of both these angles can also be inferred geometrically from Fig. \ref{fig:1}(b) ($\theta(t) < \theta(t+1), \gamma(t) > \gamma(t+1))$ thereby justifying the trend.
It is also inferred that the true and estimated values of the elevation angle are quite comparable, thus showing efficacy of the proposed filtering framework.  

It may be noted that negative elevation angles correspond to the coordinates when the LEO satellite is out of visibility zone (shown in Fig. \ref{fig:2}(a)). On the contrary, positive elevation angles correspond to the time instants where the satellite is visible to the UE, and communication is possible. 


Through Fig. \ref{fig:2}(c) we illustrate the variation of Doppler shift experienced by the UE. We study the system for two downlink frequencies of interest, $10.9 \mbox{GHz}$ and $28 \mbox{GHz}$. In general it is observed when the LEO satellite network is operated at higher transmit frequency, the UE experiences a higher Doppler shift (in range of 680 KHz at 28 GHz in contrast to around 260 KHz at 10.9 GHz). Furthermore, the Doppler shift is observed to decrease with increasing elevation angle. This can be attributed to the fact that larger elevation angles correspond to closer slant range between the satellite and UE, thereby reducing the Doppler shift experienced.

In Figs. \ref{fig:3}(a)-(b), we illustrate the variation of timing advance first with time and then with elevation angles. The TA is observed to vary from around $[10 \mbox{ms}, 90 \mbox{ms}]$ when the satellite is visible to the UE. 
It is inferred (from Fig. \ref{fig:3}(b)) that the TA value attains the minima at the maximum elevation angle, and the maxima corresponding to the minimum elevation angle. This trend can be attributed to the fact that the $\theta_{max}$ corresponds to the  minimum slant range between the UE and the satellite, thereby resulting in minimum TA. Additionally, the TA value variation is also proportional with the variation of the radial distance of the LEO satellite (from the earth center) as can be inferred from Fig. \ref{fig:3}(a). 

The proposed framework is observed to be extremely robust and efficient to study and estimate key communication parameters like TA, doppler shift, and satellite visibility at a local area on earth surface, which are paramount to establish a LEO satellite communication network.



\section{Conclusion}\label{SecVI}
This work has presented a robust and accurate framework to estimate the timing advance and Doppler shift experienced at the UE in a LEO satellite based non-terrestrial network. The framework has captured the joint system dynamics through an EKF based recursive Bayesian framework. The LEO satellite motion model has been accurately modeled by considering the satellite acceleration. The Jacobian corresponding to the proposed motion model has also been constructed to linearize the system non-linearities. Probabilistic insights regarding the EKF based filtering process has also been discussed. The framework computes the TA and Doppler shift by taking into account the limited LEO satellite visibility and the LEO-UE geometry with respect to the earth center. The results have illustrated the robustness and efficacy of the proposed estimation framework especially at high UE velocities.The proposed framework is expected to be extremely useful for realizing upcoming LEO satellite based non-terrestrial networks.




\end{document}